\documentclass[%
 12pt,
 reprint,
superscriptaddress,
 amsmath,amssymb,
 aps,
pra,
floatfix,
longbibliography
]{revtex4-2}

\usepackage{microtype}
\usepackage{graphicx}% Include figure files
\usepackage{dcolumn}% Align table columns on decimal point
\usepackage{bm}% bold math

\usepackage{xcolor}
\usepackage{booktabs}
\usepackage{multirow}

\usepackage{lipsum}
\usepackage{physics}

\usepackage{siunitx}
\usepackage{hyperref}% add hypertext capabilities
\hypersetup{%
  colorlinks,
  linkcolor={red!50!black},
  citecolor={blue!50!black},
  urlcolor={blue!80!black}
  }
\usepackage[capitalise]{cleveref}
\usepackage{url}

\usepackage{placeins}

\usepackage{cleveref}

\newcommand{\CNOT}{{\ensuremath{\rm CNOT}}}
\newcommand{\SWAP}{{\ensuremath{\rm SWAP}}}
\newcommand{\Had}{{\ensuremath{\rm H}}}
\DeclareMathOperator{\Beta}{Beta}

\begin{document}

\title{Demonstration of logical qubits and repeated error correction\\
  with better-than-physical error rates}

\author{A.~Paetznick}
\affiliation{%
Microsoft Azure Quantum
}%

\author{M.~P.~da~Silva}
\affiliation{%
Microsoft Azure Quantum
}%

\author{C.~Ryan-Anderson}
\affiliation{%
Quantinuum
}%

\author{J.~M.~Bello-Rivas}
\affiliation{%
Microsoft Azure Quantum
}%

\author{J.~P.~Campora III}
\affiliation{%
  Quantinuum
}%

\author{A.~Chernoguzov}
\affiliation{%
  Quantinuum
}%

\author{J.~M.~Dreiling}
\affiliation{%
  Quantinuum
}%

\author{C.~Foltz}
\affiliation{%
  Quantinuum
}%

\author{F.~Frachon}
\affiliation{%
Microsoft Azure Quantum
}%

\author{J.~P.~Gaebler}
\affiliation{%
  Quantinuum
}%

\author{T.~M.~Gatterman}
\affiliation{%
  Quantinuum
}%

\author{L.~Grans-Samuelsson}
\affiliation{%
Microsoft Azure Quantum
}%

\author{D.~Gresh}
\affiliation{%
Quantinuum
}%

\author{D.~Hayes}
\affiliation{%
  Quantinuum
}%

\author{N.~Hewitt}
\affiliation{%
  Quantinuum
}%

\author{C.~Holliman}
\affiliation{%
  Quantinuum
}%

\author{C.~V.~Horst}
\affiliation{%
  Quantinuum
}%

\author{J.~Johansen}
\affiliation{%
  Quantinuum
}%

\author{D.~Lucchetti}
\affiliation{%
  Quantinuum
}%

\author{Y.~Matsuoka}
\affiliation{%
  Quantinuum
}%

\author{M.~Mills}
\affiliation{%
  Quantinuum
}%

\author{S.~A.~Moses}
\affiliation{%
  Quantinuum
}%

\author{B.~Neyenhuis}
\affiliation{%
  Quantinuum
}%

\author{A.~Paz}
\affiliation{%
Microsoft Azure Quantum
}%

\author{J.~Pino}
\affiliation{%
  Quantinuum
}%

\author{P.~Siegfried}
\affiliation{%
  Quantinuum
}%

\author{A.~Sundaram}
\affiliation{%
  Microsoft Azure Quantum
}%

\author{D.~Tom}
\affiliation{%
Microsoft Azure Quantum
}%

\author{S.~J.~Wernli}
\affiliation{%
Microsoft Azure Quantum
}%

\author{M.~Zanner}
\affiliation{%
Microsoft Azure Quantum
}%

\author{R.~P.~Stutz}
\affiliation{%
  Quantinuum
}%

\author{K.~M.~Svore}
\affiliation{%
  Microsoft Azure Quantum
}%

\date{\today}

\begin{abstract}
  The promise of quantum computers hinges on the ability to scale to
  large system sizes, e.g., to run quantum computations consisting of
  more than 100 million operations fault-tolerantly. This in turn
  requires suppressing errors to levels inversely proportional to the
  size of the computation.  As a step towards this ambitious goal, we
  present experiments on a trapped-ion QCCD processor where, through
  the use of fault-tolerant encoding and error correction, we are able
  to suppress logical error rates to levels below the physical error
  rates. In particular, we entangled logical qubits encoded in the
  $[[7,1,3]]$ code with error rates $9.8\times$ to $500\times$ lower
  than at the physical level, and entangled logical qubits encoded in
  a $[[12,2,4]]$ code based on Knill's C4/C6 scheme with error rates
  $4.7\times$ to $800\times$ lower than at the physical level,
  depending on the judicious use of post-selection.  Moreover, we
  demonstrate repeated error correction with the $[[12,2,4]]$ code,
  with logical error rates below physical circuit baselines
  corresponding to repeated \CNOT s, and show evidence that the error
  rate per error correction cycle, which consists of over 100 physical
  \CNOT s, approaches the error rate of two physical \CNOT s. These
  results signify a transition from noisy intermediate
  scale quantum computing to reliable quantum computing, and
  demonstrate advanced capabilities toward large-scale fault-tolerant
  quantum computing.
\end{abstract}

\maketitle

Quantum computers have the potential to solve important
classically-intractable problems, however doing so requires improving
error rates well beyond those of the underlying physical hardware. The
development of quantum error correction and fault-tolerant quantum
computing was a major theoretical breakthrough that paved the way for
the implementation of reliable quantum
computers~\cite{ShorQEC1995,Steane1996,Dorit1997,kitaev1997quantum,Knill1998,Terhal2005,Aliferis2006}. Without
quantum fault-tolerance, there is little to no indication that quantum
computers can solve important practical problems that are outside the
reach of modern day supercomputers and machine
learning~\cite{Reiher2017,Beverland2022}. The experimental challenges
remain significant, as fault tolerance requires that physical error
rates be sufficiently low before the overhead of error correction
leads to an improvement over physical, non-fault-tolerant
operations~\cite{Dorit1997,Knill1998,Terhal2005,Aliferis2006,Raussendorf2007}.
Steady progress has been made across several different
platforms. Several experiments have shown indications of
physical error rates approaching this important so-called
threshold~\cite{egan2020fault,RyanAnderson2021,Acharya2023,Sivak2023,RyanAnderson2022},
while others have demonstrated operations on multiple logical
qubits~\cite{Erhard2021,Postler2021,Bluvstein2023}. 
However, to the
best of our knowledge none of these experiments have convincingly 
demonstrated error correction with logical error rates better than the physical error rates.
\footnote{A notable exception being a demonstration of Bell correlations that are stronger than physical correlations~\cite{RyanAnderson2022}. Since the original writing, other notable demonstrations have
occured, including definitive sub-threshold
performance~\cite{acharya2024belowthreshold}, and additional
better-than-physical performance~\cite{reichardt2024dtesseract}.}

Our goal is to demonstrate the transition from noisy intermediate
scale quantum computing to reliable quantum
computing~\cite{Svore2023,Haah2024}, through the co-optimization of
hardware and software with a present-day commercial quantum processor.
Namely, we aim (1) to show convincingly a large separation between
logical and physical error rates, (2) in a setting where {\em all
single circuit faults are corrected}, while (3) using logical circuits
representative of what would be used for computation, e.g., the
preparation or use of logical entanglement.

To this end, we demonstrate several fault-tolerant protocols in a
commercial trapped-ion quantum charge-coupled device (QCCD)
processor~\cite{Moses2023} and show that the observed logical error
rates are conclusively lower than the error rates for their
(unencoded) physical counterparts.

\section{Methodology\label{sec:methodology}}

Our approach largely builds on Gottesman's proposal in
Ref.~\onlinecite{Gottesman2016}. Namely, we benchmark {\em complete
quantum circuits}~\footnote{Complete quantum circuits are circuits
that prepare qubits in a fixed state, perform a sequence of gates, and
measure one or more qubits to yield classical output bits.}, and
contrast the error rates of the classical outputs of these
circuits---a comparison between the outputs of the unencoded physical
circuit to that of the corresponding fault-tolerantly encoded circuit
on the same hardware. We deviate from Gottesman's proposal in two ways
-- we consider a different metric for the comparison of quantum
circuits, and we allow for slightly more general state preparations.

The metric considered in Gottesman's proposal is the {\em total
variation distance} between the output distribution of the ideal
circuit and the experimental circuits (encoded or otherwise). For our
proposal we also consider the statistical distance between outputs,
but we add classical processing of the measurement outputs so that it
is possible to determine success or failure for each individual run of
the experiment. For example, when preparing the state
$\frac{\ket{00}}{\sqrt{2}}+\frac{\ket{11}}{\sqrt{2}}$ and measuring
both qubits in either the $Z$ or the $X$ basis at random, we consider
success if both outcomes agree and failure if they disagree. Due to
the additional classical processing (i.e., comparing the two bits and
reporting only their parity) this metric is generally weaker than the
distance proposed by Gottesman, but it simplifies the analysis and
yields estimates with much lower uncertainty (an important
consideration in an experimental setting with finite
resources). Moreover, in many cases of interest, experiments of this
type allow for direct estimates of the state fidelity through the
measurement of the parity of the stabilizer group
elements~\cite{Flammia2011,Silva2011}.

We also deviate from Gottesman's proposal by allowing that the qubits
be prepared in some finite set of fixed states instead of only the
$\ket{0}$ state, since it is natural to consider specialized
preparations of resource states in the encoded setting. In the
unencoded setting the preparation will reduce to preparation of qubits
in the $\ket{0}$ state and application of various gates.

% section: HW and compiler details
\section{Hardware platform\label{sec:hw-compiler-details}}

All reported demonstrations here were performed on Quantinuum's
H2 trapped-ion processor. The device was recently reported on in
detail in Ref.~\onlinecite{Moses2023}, and we give a brief overview here.

H2 is a shuttling-based trapped-ion QCCD
device~\cite{Wineland98,Pino2020}. H2 has all the necessary
ingredients for state-of-the-art quantum error correction experiments:
high-fidelity state-preparation and measurement (SPAM) with an error rate of
$0.15\%$, two-qubit gates with an error rate of $0.14(1)\%$, long
range connectivity (a key enabler for running some of the circuits
described here), and mid-circuit measurement and reset with crosstalk
errors $\leq2\times10^{-5}$.

The H2 processor is commercially available through both
Quantinuum~\cite{quantinuum} and Azure Quantum~\cite{azure-quantum}.
All experiments in
\cref{subsec:carbon-code-bell,sec:repeated-error-correction} were
submitted through the Azure Quantum software stack using the Quantum
Intermediate Representation (QIR)~\cite{qir-alliance}.  For the
experiments described in \cref{sec:repeated-error-correction},
additional compiler customizations were applied, as described in that
section in more detail.

% section: Bell state preparation
\section{Improved logical entanglement}

Entanglement is a quantum hallmark, and over the last 20 years the
preparation and measurement of Bell states has been a baseline
demonstration for credible physical implementations of quantum
computers.  It is natural to consider the same baseline in the context
of logical qubits~\cite{ELQ-IARPA2024}, as the complexity of the
fault-tolerant preparation already becomes apparent.  Several recent
experiments have demonstrated the preparation of logical Bell-pairs in
quantum error correction
codes~\cite{Postler2021,RyanAnderson2022,Bluvstein2023}, and although
these results are remarkable in their own right, only one of
them~\cite{RyanAnderson2022} has demonstrated a logical error rate
modestly better than physical error rate. Here we describe how much
lower logical error rates can be obtained using the $[[7,1,3]]$ Steane
code and a $[[12,2,4]]$ code.

% subsection: Steane code Bell prep
% Bell state preparation
\subsection{Steane code\label{subsec:steane-code-bell}}

We first present results using the $[[7,1,3]]$ Steane code~\cite{Steane1996} to prepare a high-fidelity Bell state. The Steane code, or distance-three color code, has been used in several demonstrations of logically encoded circuits~\cite{nigg2014quantum,hilder2021faulttolerant,Postler2021,RyanAnderson2021,RyanAnderson2022,Bluvstein2023}, partially because of its relatively low space-time overhead, its simple preparation and measurement protocols due to its CSS nature, and all single and two qubit Clifford gates being transversal for the code.

\subsubsection{Circuits}

\begin{figure}
\includegraphics[scale=1.5]{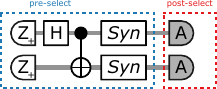}
\caption{High-level depiction of the logical program of the Bell
  resource-state preparation using the Steane code. The blue dashed
  box indicates the pre-selected portion of the circuit where both the
  verification of $\ket{0}$ and trivial measurement results of
  syndrome extraction rounds (the boxes labeled $Syn$) are used to
  verify the creation of a resource state. The experiments are
  analyzed using error correction (with pre-selection) or error
  detection (with pre- and post-selection)
  independently. Post-selection accepts experiments where the syndrome
  inferred from the logical measurements (red dashed boxes) is
  trivial.\label{fig:steane-bell}}
\end{figure}

The circuit components used to generate a high-fidelity Bell state were previously demonstrated in Refs.~\onlinecite{RyanAnderson2021,RyanAnderson2022}. The logical program to prepare a logical Bell resource state using the Steane code is in Fig.~\ref{fig:steane-bell}. The preparation includes encoding circuits to initialize two logical qubits to $\ket{0}$, transversal single and two-qubit Clifford gates, flagged syndrome extraction, and destructive logical measurements. Each logical qubit has seven data qubits and three ancilla qubits, leading to the experiments having a total of 20 physical qubits.

The encoding circuit is made fault-tolerant using a scheme by Goto~\cite{goto2016minimizing} which involves a non-fault tolerant encoding circuit of $\ket{0}$ followed by a verification step where an ancilla qubit measures the logical $Z$ operator. Upon failure to verify the preparation a logical $\ket{0}$ state, the qubits can be conditionally reset and the fault-tolerant preparation can be re-attempted in a repeat-until-success fashion (as seen in Ref.~\onlinecite{RyanAnderson2021}) or pre-selected upon verification. For these experiments, we chose to repeat the preparation of the $\ket{0}$ up to three times. Once the $\ket{0}$ states are verified, the Goto scheme ensures that state preparation results in at most weight-one faults due to any single faulty gate. 

After both logical qubits are prepared in $\ket{0}$, we apply a transversal Hadamard to one of the logical qubits, followed by a transversal \CNOT~between the two logical qubits. Ideally, this circuit would produce $(\ket{00}+\ket{11})/\sqrt{2}$. 

We attempt to verify this by following the \CNOT~with one round of flagged syndrome extraction on each logical qubit based on a scheme by Chao and Reichardt~\cite{chao2018fault} in which three syndromes are measured in parallel. The main difference compared to the original scheme is that we do not follow the flagged syndrome measurements with a conditional set of unflagged syndrome measurements. This is because we are treating the Bell state as a fixed resource state independent of the computation, and much like the the logical $\ket{0}$, pre-selection of such resources has no negative impact on scalability~\cite{ShorQEC1995,preskill1998reliable,bravyi2005universal}.

After running syndrome extraction, logical single-qubit transversal gates are applied to measure in appropriate logical Pauli bases, and then we destructively measure the data qubits. We measure both qubits in the logical $X$, $Y$, and $Z$ bases, which can be performed transversally in the Steane code. The destructive measurements of data qubits not only allows one to determine logical outcomes but can also be used to determine syndromes. At the end of each destructive logical measurement, these syndromes are used to generate a correction to the logical outcome using a lookup table decoder. These corrected outcomes are determined by running the decoder during the hybrid quantum/classical program on the device and not determined afterwards. The lookup table is relatively simple and the same one used in Ref.~\onlinecite{RyanAnderson2021}. Syndromes are decoded independently for each logical measurement. Further improvements could be made by incorporating experimental bias noise in the construction of the decoder, as well as decoding over all syndromes generated by both logical qubits~\cite{Bacon2017,Gottesman2022,Delfosse2023,Bluvstein2023}.

The $[[12,2,4]]$ code also studied in this work does not admit for a
transversal fault-tolerant $Y$ eigenbasis measurement, so to
facilitate comparison we focus on the error rate for Bell experiments
where we measure only $X$ and $Z$ parities, and denote it
$E_{xz}$. Simply put, $E_{xz}$ is defined as the number of ``incorrect'' measurement results divided by the total number of trials. Only measuring in the $X$ and $Z$ basis is related to measuring the stabilizers of the target state. These measurements can provide a bound on the state fidelity, but since we can also measure in the $Y$ basis when using the Steane code, we also report estimates of the state fidelity in~\cref{app:steane-data}. The state fidelities give a 
more conservative estimate of the overall performance of the Bell preparation protocol; however,
the state fidelities paint a similar picture to the estimates given by $E_{xz}$, where the encoded circuits are significantly out performing the results of the unencoded circuit analog.

Besides evaluating $E_{xz}$ in the presence of quantum error
correction, we also re-analyze the corrected outcomes calculated by the 
device by excluding any results for which non-trivial syndromes were recorded 
during the destructive measurement phase and the evaluate the Steane code as 
a quantum error detecting (QED) code. This mode of operation does not satisfy 
the requirement that all single faults must be corrected, but allows us to
evaluate the probability of the environment applying pure logical errors
(thus serving as an upper bound on QEC performance) as well as evaluate the 
rejection rate cost of using the Steane code in a QED manner and whether 
that cost is worth paying to further suppress noise.  

As described in \cref{sec:methodology}, we compare $E_{xz}$ of
physical (unencoded) circuits and logical circuits. The physical circuits 
are identical to what is described in
\cref{fig:steane-bell}, with the exception that no syndrome
extraction, pre-selection, or post-selection is applied.  Since H2 has
four gate zones, four copies of the physical circuits were executed in
parallel for each run, totaling eight physical qubits per program.

\subsubsection{Experimental results}

\begin{figure}
    \includegraphics[width=\columnwidth]{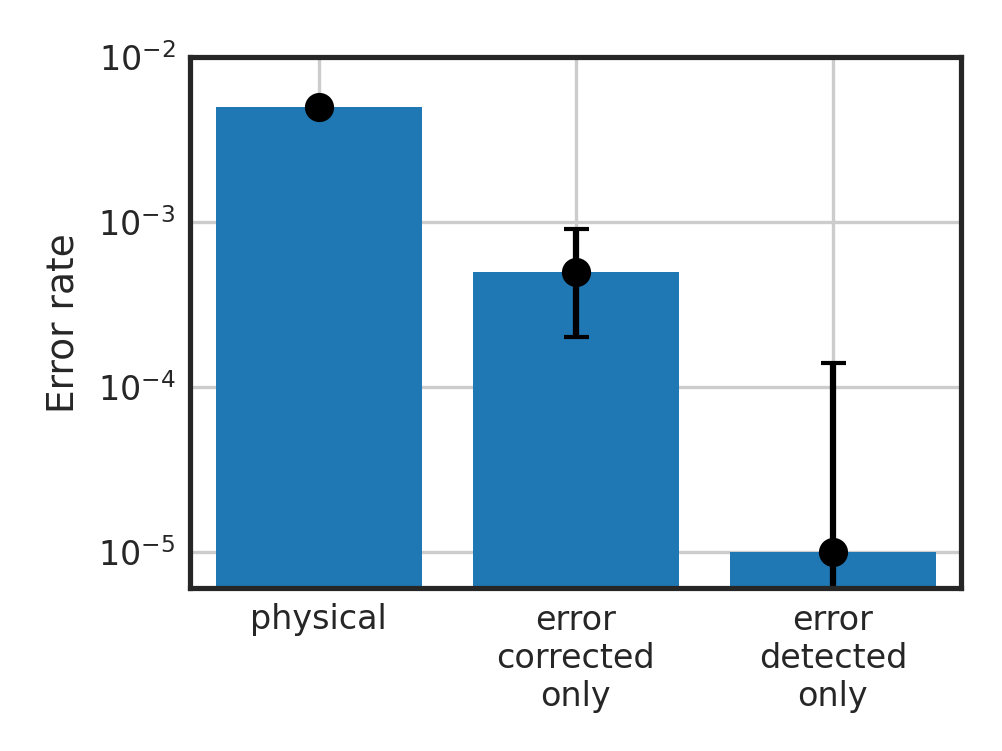}
    \caption{Results for the Bell state preparation in the $[[7,1,3]]$
      Steane code comparing physical level, error correction, and error
      detection experiments. There is a statistically significant
      separation between the physical and encoded results given the
      errors bars, which indicate $95\%$ confidence
      intervals. \label{fig:steane-bar}}
\end{figure}

The experimental results for the both the Steane code and physical
level Bell state preparation are summarized in
Table~\ref{Table:BellSteaneSum} and Fig.~\ref{fig:steane-bar} (see
\cref{app:statistical-analysis} for details of the statistical
analysis and \cref{app:steane-data} for additional data).

We ran a total of $411,600$ unencoded experiments (four Bell state preparation and measurement circuits per program). When restricted to $X$ and $Z$ basis experiments, $274,400$ experiments were ran at the unencoded level, respectively. For both the unencoded and encoded set of experiments, we ran an equal number of sub-experiments measuring $X$, $Y$, and $Z$ correlations. Of the $411,600$ physical experiments, $1,897$ measured the incorrect parity. In particular, the wrong parity was measured $572$ out of $137,200$ times for $X$ parity, $530$ out of $137,200$ times for $Y$ parity, and $795$ out of $137,200$ times for $Z$ parity. This results in an error rate $E_{xz}=0.50\% _{-0.03\%}^{+0.03\%}$. Looking at the outcomes for separate bases (see Table~\ref{Table:BellSteanePhys}), it is apparent that the $Z$ measurements experience an increased error rate of approximately $0.6\%$ compared to about $0.4\%$ for the $X$ and $Y$. This difference may arise due to biased noise in the two-qubit gates.

For the logical experiments with error correction and pre-selection,
$12,100$ experiments were ran for each of $X$, $Y$,
and $Z$ correlations, for a total of $36,300$
experiments.  About $9,000$ shots were pre-accepted for each basis
resulting in a pre-acceptance rate of about $75\%$. Out of the $9,025$
pre-accepted $X$ correlations experiments, $337$ non-trivial syndromes
were measured and $9$ experiments resulted in the wrong parity
measurement. For the $9,082$ pre-accepted $Y$ correlation experiments,
$417$ non-trivial syndromes were measured and $8$ incorrect parity
measurements were made. For $Z$ correlations, of the $9,010$ pre-accepted
experiments, $309$ had non-trivial syndromes and $0$ measurements
resulted in measuring the wrong parity.

Post-selecting on non-trivial syndromes (so no error correction is
performed, only error detection), the total acceptance rate goes from
$\approx75\%$ to $\approx72\%$, and $0$ incorrect parity outcomes are
observed for the $X$ and $Z$ correlations out of $8,688$ and $8,701$
post-accepted experiments, respectively. For the $Y$, only $2$
measurements out of $8,665$ experiments accepted in post-selection
(post-accepted) resulted in a measurement of the wrong parity.

Based on these observations, the error rates $E_{xz}$ were determined
to be $0.50\% _{-0.03\%}^{+0.03\%}$, $0.05 \% _{-0.03\%}^{+0.04\%}$,
and $0.001 \% _{-0.001\%}^{+0.013\%}$ for the physical experiments,
logical experiments with error correction, and logical experiments
with error detection, respectively, as illustrated in
\cref{fig:steane-bar}. This corresponds to a statistically
significant reduction in the observed error rate by factors of $9.8$
and $500$ for the error corrected and the error detected experiments
respectively.

% Steane Summary
\begin{table*}[t]
  \centering
  \renewcommand{\arraystretch}{1.5}
\begin{tabular}{lrrrrrlr}
\toprule
           & runs & pre-accepted & post-accepted &  corrections & errors & $E_{xz}$ error rate & gain \\ 
\hline
unencoded baseline    & $274,400$ & --- & --- & --- & $1,367$ & $0.50\% _{-0.03\%}^{+0.03\%}$      &  ---     \\ 
encoded, pre-selection only   & $24,200$ & $18,035$  & --- & $646$ & $9$  & $0.05 \% _{-0.03\%}^{+0.04\%}$     & $9.8$   \\  
encoded, pre- and post-selection & $24,200$ & $18,035$ & $17,389$ & --- & $0$ & $0.001 \% _{-0.001\%}^{+0.013\%}$     & $500$ \\ 
\bottomrule
\end{tabular}
\caption{Summary of experimental results for the preparation of Bell
  resource states using the $[[7,1,3]]$ Steane code and the
  corresponding circuit composed of physical qubits instead of logical
  qubits. Note this table only includes data for the $X$ and $Z$ basis
  measurements. Additional data including the estimated state
  fidelity are given in Appendix~\ref{app:steane-data}. We define
  ``gain'' to be the error rate of the physical circuits divided by
  the error rate of the corresponding logical circuit, while
  ``corrections'' refers to the number of pre-accepted experiments where a
  non-trivial syndrome was observed.}
\label{Table:BellSteaneSum}
\end{table*}

% subsection: Carbon code Bell prep
\subsection{Carbon code\label{subsec:carbon-code-bell}} 

\begin{figure}
  \includegraphics[scale=1.5]{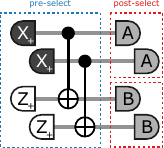}
  \caption{Effective logical circuit for Bell state preparation using
  the $[[12,2,4]]$ Carbon code. The top and bottom pair of lines
  correspond to separate blocks.  The pre-selected portion of the
  circuit (blue dashed box) include verification measurements at the
  physical level which are used for pre-selection (not shown). When
  post-selection is performed, it is based only on the syndrome
  information in the transversal measurements (red dashed boxes), and
  a separate decision is made for each code block. The logical
  observables $A$ and $B$ can be either $X$ or $Z$ independently, but
  for the experiments discussed here we focus on the scenario where
  $A=B$.
  \label{fig:carbon-bell-circuit}}
\end{figure}

We also entangled two pairs of logical qubits using a $[[12,2,4]]$ CSS
code, which we nickname {\em Carbon}~(see 
\cref{app:carbon} for details), and perform Bell correlation
experiments with this entangled state. Carbon is the result
of concatenating the $[[4,2,2]]$ code with the $[[6,2,2]]$ code, as
proposed originally by Knill~\cite{Knill2003,Knill2005}.  The main
modification of Knill's original proposal is that the circuits here do not
obey the concatenated structure, resulting in much
more compact state preparation and syndrome extraction circuits.  As a CSS code,
Carbon has transversal \CNOT\ and \Had\ gates, which immediately
enable the preparation of a maximally entangled logical Bell
state. However, the transversal \CNOT\ gates create entanglement {\em
  between} qubit blocks, instead of entanglement {\em within} the
block. For that reason, the Bell correlation experiment now involves
two logical Bell states, as illustrated in
\cref{fig:carbon-bell-circuit}.

\subsubsection{Circuits}

The Carbon code circuit for logical Bell state preparation requires 30
physical qubits (24 data qubits for the two blocks and 6
ancillas). While it is possible to fault-tolerantly prepare each block
in tensor products of $X$ or $Z$ eigenstates and then prepare Bell
states by applying the transversal \CNOT\ between them, it is more
favorable to distribute the verification measurements throughout the
Bell state preparation circuit. See~\cref{fig:c12-00,fig:c12-bell}. These verification
circuits are tailored to the Bell state preparation, and can avoid
measuring all stabilizer generators, focusing instead of simply
detecting the propagation of failures into high weight Pauli
errors. This simplification improves the fidelity of accepted state
preparations in the experiments at the cost of mildly increasing the
rejection rate.

We emphasize that the pre-selection criteria only looks to detect
correlated error propagation in the unitary preparation circuits, as
is customary in fault-tolerant stabilizer state preparation
circuits. Most importantly, we do not measure the Bell correlations
non-destructively in the preparation circuit, nor do we measure all
stabilizer generators for the state.

The $X$ and $Z$ parities for each Bell pair can be obtained by
measuring each block transversally, measuring both qubits within each
block in the same basis~\footnote{The measurement of $Y$ parities
requires more complex circuitry (effectively applying the $S$ gate to
change bases), so we leave these more complex experiments for future
work.}.  As in \cref{subsec:steane-code-bell}, no error correction is
performed between state preparation and measurement, but syndrome
information from the transversal measurements allows us to detect
errors nonetheless. Because the Carbon code has distance 4, while we
can correct all weight 1 errors in each block, we cannot correct all
weight 2 errors. However, we can {\em detect} all errors of weight up
to 2 and reject them, if necessary.
Thus, the Carbon code
experiments can have two modes of operation: an {\em error correction}
mode, and an {\em error correction and rejection} mode. 
Error correction is performed in both modes of operation.

In error correction mode, we only perform pre-selection of state
preparation, and do not allow any post-selection based on the syndrome
information. Every experimental run that is accepted in pre-selection
(or {\em pre-accepted}) yields logical experimental outcomes that may
or may not be corrected based on syndrome information. Decoding is
performed by table lookup after collecting the experimental data, 
and since we assume errors are independent
and identically distributed, the table is populated sequentially by
examining Pauli errors of increasing weight. If two Pauli errors of a
particular weight have the same syndrome and their product is not in
the stabilizer group, we choose arbitrarily which correction
we associate with that syndrome, knowing that, for that particular
syndrome we only make the right correction a constant fraction of the
time. These resulting logical outcomes are then categorized as
failures or successes based on the expected outcomes of a noiseless
circuit (i.e., having parity outcome $+1$ for each Bell pair).

In the second mode of operation, we perform pre-selection of state
preparation as before, but we also {\em post-select} on syndrome
outcomes. Post-selection is based on the expected quality of decoding
decisions. We modify the decoder table construction 
so that pairs of circuit faults are no longer accepted.
Details are discussed in \cref{app:carbon}.

The advantage of post-selecting on some syndrome outcomes for even
distance codes is that it effectively boosts the distance of the code
from $d$ to $d+1$, in the sense that errors of weight $\frac{d}{2}$ do
not cause logical errors anymore (as those are rejected in
post-selection)~\cite{Prabhu2021}. Although post-selection is not
scalable in a strict sense, with low rejection rates one can run
relatively deep circuits and achieve significant improvements to the
logical error rate~\cite{Prabhu2021,Chen2022}. The key observation is
that the post-rejection rate is second order in the physical error
rate, since single faults are corrected instead of rejected.

For the experiments here, the pre-rejection rates are a limiting
factor, since the pre-rejection rate is first order in the physical
error rate. However, with state preparation factories in a larger
system, pre-selection is scalable, and post-selection becomes the
dominant source of overhead.

\subsubsection{Experimental results}

We ran a total of 16,000 unencoded experiments, and 22,000 encoded
experiments~\footnote{We also ran experiments where we measured $XZ$
and $ZX$ cross-parities for each Bell pair, and confirmed that the
distribution was close to uniform.}. In both cases half of the runs
measured $X$ parities, and half of the runs measured $Z$
parities.

For the unencoded experiments, out of the 16,000 runs, 125 yielded the
incorrect parity, resulting in a physical error rate of
$0.8\%_{-0.1\%}^{+0.1\%}$.

\begin{figure}
    \includegraphics[width=\columnwidth]{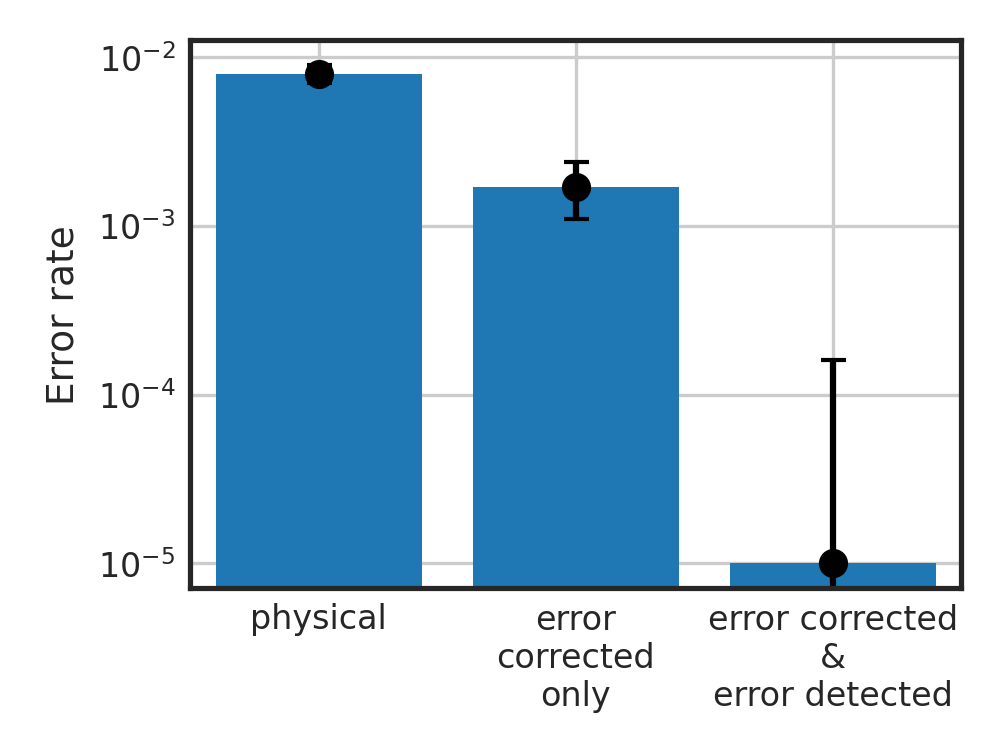}
    \caption{Comparison between physical and logical error rates of
      the $[[12,2,4]]$ Carbon code for the Bell state preparation
      circuit in \cref{fig:carbon-bell-circuit}. Precise numerical
      values can be found in \cref{tab:bell-carbon-results}. The
      difference in the error rates of the physical versus logical
      levels is statistically significant, as illustrated by the
      separation between the $95\%$ confidence intervals.
      \label{fig:bell-carbon-results}}
\end{figure}

For the encoded experiments, the pre-selection procedure reduced the
data set to a total of 15,483 out of the 22,000 encoded
runs, roughly equally distributed across $X$ and $Z$ parities (a
pre-acceptance rate of roughly $70\%$). Out of the 15,483 pre-accepted
runs, 928 had a non-trivial syndrome, triggering a correction from the
decoder, and in 26 experiments the resulting parity outcome was
incorrect. When we allow for post-selection, the total number of
accepted runs is reduced to 15,409, and 854 of these runs trigger a
correction by the decoder (74 of the runs that triggered ambiguous
syndromes were rejected by post-selection), and the logical parity was
correct in all accepted runs. Note that the post-rejection rate is
roughly $0.5\%$, two orders of magnitude lower than the pre-rejection
rate, matching our expectations. These experimental results are summarized
in \cref{tab:bell-carbon-results} and \cref{fig:bell-carbon-results}.

Using the methods described in \cref{app:statistical-analysis}, we
estimate the error rate for the unencoded circuit is
$0.8\%_{-0.1\%}^{+0.1\%}$, while our estimate for the encoded
experiments with pre-selection is $0.17\%_{-0.06\%}^{+0.07\%}$ (a
reduction in error rate by a factor of 4.7). Our estimate
for the encoded experiments with pre-selection and post-selection is
$0.001\%_{-0.001\%}^{+0.015\%}$, corresponding to a reduction in error
rate by a factor of 800.

\begin{table*}[t]
  \centering
  \renewcommand{\arraystretch}{1.5}
\begin{tabular}{lrrrrrlr}
    \toprule
    & runs   & pre-accepted & post-accepted & corrections & errors & $E_{xz}$ error rate & gain \\
    \hline
unencoded baseline  
& \num{16000} &  \num{16000} & ---           & ---         &    125 & $0.8\substack{+0.1\\-0.1}\%$     & ---\\
encoded, 
pre-selection only 
& \num{22000} &  \num{15483} & ---           & 928         &     26 & $0.17\substack{+0.07\\-0.06}\%$     & 4.7\\
encoded, pre- 
and 
post-selection     
& \num{22000} &  \num{15483} & \num{15409}   & 854         &      0 & $0.001\%_{-0.001\%}^{+0.015\%}$     & 800 \\
    \bottomrule
\end{tabular}
\caption{Summary of experimental results from Bell correlation
  experiments with the $[[12,2,4]]$ Carbon code. We define ``gain'' to
  be the error rate of the unencoded circuits divided by the error
  rate of the encoded circuit in question, while ``corrections''
  refers to how often a non-trivial syndrome was observed in the
  experiments that were accepted and pre- and/or
  post-selection.\label{tab:bell-carbon-results}}
\end{table*}

% section: Repeated error correction
\section{Repeated fault-tolerant error correction\label{sec:repeated-error-correction}}

It is not sufficient to show shallow logical circuits that outperform
their physical counterparts. The promise of quantum computers lies
with solving large practical problems that require deep quantum
circuits~\cite{Reiher2017,Beverland2022}. This, in turn, requires
fault-tolerant gadgets that intermix quantum error correction with
fault-tolerant logical
operations~\cite{Dorit1997,Knill1998,Terhal2005,Aliferis2006}. Composable,
repeatable error correction is one of the key milestones in this
journey, and only recently experimental demonstrations have become
possible~\cite{Andersen2020,Sundaresan2023,Acharya2023}. Two 
demonstrations from the last several years deserve special mention.
A demonstration for the $[[7,1,3]]$ Steane code achieved
$\approx1.75\%$ logical error rate per error correction cycle for up
to 6 rounds of error correction~\cite{RyanAnderson2021}, a logical
error rate just under one order of magnitude higher than the dominating
error rate for the physical operations.  Repeated error correction was
also demonstrated for the surface code~\cite{Acharya2023}, achieving
error rates of $\approx9\%$ to $\approx15\%$ per error correction
cycle for up to 8 rounds of error correction, which is again just
under one order of magnitude higher than the dominating error rate for
the physical operations.
\footnote{We again note demonstrations~\cite{acharya2024belowthreshold,reichardt2024dtesseract} 
that appeared since our original writing.}

Here we demonstrate as many as 3 rounds of error correction for the
$[[12,2,4]]$ Carbon code, using a combination of error correction and
detection. The Carbon code has a high threshold of approximately $3\%$
and a rate of $\frac{1}{6}$ at distance 4 (see~\cref{app:carbon}), which
makes it a good candidate for experimental demonstrations.
We compare the error rates of logical circuits with error
correction and physical circuits, and show small circuits with logical
error rates lower than the physical error rates by a statistically
significant margin. We also give evidence that the error rate
accumulated per round of error correction is comparable to the error
rate accumulated with two physical \CNOT\ in series.

\subsection{Carbon code circuits}

In the spirit of Ref.~\onlinecite{Gottesman2016}, instead of simply
showing individual logical circuits that outperform their physical
counterparts, we would like to show that families of circuits composed
to perform some computation benefit from encoding. A crucial element
of such a demonstration is repeated fault-tolerant error correction.

Despite the appealing features of the Steane code (such as transversal
Clifford group and relatively compact state preparation and syndrome
extraction circuits), estimates for the error threshold of the code
remain relatively low, making experimental demonstrations of logical
circuits outperforming physical circuits challenging.

The Carbon code, on the other hand, has a high threshold under the
usual assumptions about parallelism, access to fresh ancillas,
uniformity of the noise~\cite{Knill2003,Knill2005}. This proposal relies on
teleportation-based syndrome extraction to achieve this
performance~\cite{Knill2003}. Such an approach
requires access to 3 code blocks within the hardware (one block for
the data, and two blocks for the maximally entangled resource state).

\begin{figure*}
  \includegraphics[scale=1]{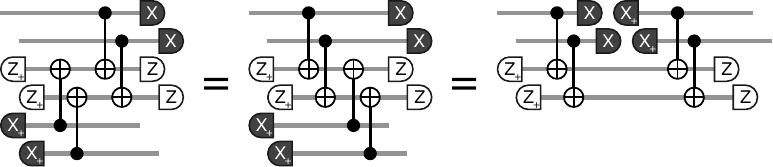}
  \caption{Syndrome information can be obtained by performing
    teleportation at the logical level, as described by
    Knill~\cite{Knill2003}. Taking the realization of the original
    teleportation circuit for two logical qubits encoded in a Carbon
    block, which required 3 encoded blocks (left), it is possible to
    rearrange commuting circuit components to arrive at a circuit that
    uses a sequence of two 1-bit teleportations~\cite{Zhou2000} to
    extract syndrome information requiring only 2 encoded blocks at
    any given time (right)
    (see~\cref{app:carbon}).\label{fig:teleportation-syndrome-extraction}}
\end{figure*}

There is, however, a more compact option
similar to Steane syndrome extraction~\cite{Steane1997}, but derived
from Knill's approach to syndrome extraction. The key observation is
illustrated in \cref{fig:teleportation-syndrome-extraction}, resulting
in a circuit that essentially serializes the syndrome extracting
teleportation into two 1-bit teleportations~\cite{Zhou2000}---the only
requirement being that one must be able to prepare the encoded
$\ket{0}$ and $\ket{+}$ states, apply an encoded \CNOT, and measure in
the encoded $X$ and $Z$ eigenbases (and all of these operations are
available in the Carbon code).

\subsection{Physical baselines}

Within a single block, the simplest fault-tolerant operations
available for the Carbon code are the $\Had\otimes\Had$, 
and a \CNOT\---the \Had\ is implemented via transversal \Had\ application
followed by a qubit permutation, while the \CNOT\ is implemented via a
qubit permutation~\cite{Leon1982,Grassl2013}. Since qubit
permutations can be applied in the H2 system without interaction
between qubits, the permutations do not require additional ancillas to
be fault-tolerant.

\begin{figure*}
\includegraphics[scale=1]{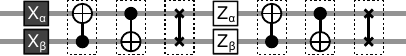}
\caption{Logical circuit family that can be implemented in a modified
  single round of error correction with the $[[12,2,4]]$ Carbon code,
  assuming
  $(\alpha,\beta,\delta,\gamma)\in\{0,\pm\frac{\pi}{2},\pm\pi\}$. The
  logical single-qubit rotations require modified state preparation
  during syndrome extraction, and the order of $X$ and $Z$ rotations
  can be transposed by modifying the order of the 1-bit
  teleportations, but neither of these modification impact the ability
  to extract syndrome information.  Any combination of gates in the
  dashed boxes can be implemented by a permutation of the
  physical qubits.
  \label{fig:1-round-logical-circuit}}
\end{figure*}

Combining gate teleportation and permutation-based gates allows for
logical circuits of the form depicted in
\cref{fig:1-round-logical-circuit} in a single round of error
correction. The \CNOT s and \SWAP s of logical qubits within a block
and their composition can be implemented by qubit permutations.

With the circuits of
\cref{fig:teleportation-syndrome-extraction,fig:1-round-logical-circuit}
in mind, we consider two physical baselines for the error correction
experiments: a sequence of two physical 1-bit teleportations, and a
sequence of two \CNOT s (since we are focused on complete
circuits~\cite{Gottesman2016}, the inputs are fixed to be a tensor
product of a pair of eigenstates of $X$ or a pair of eigenstates of $Z$,
and we choose the final measurement to be the corresponding tensor
product observable).

The comparison between physical 1-bit teleportation and logical
1-bit teleportation informs us about the improvement error correction, pre-, and
post-selection are providing over the additional
encoding overhead, but without using the rigorous fault-tolerant
gadget translation of the circuits from
Ref.~\onlinecite{Aliferis2006}. In other words, this first baseline is
a precursor for a comparison of fault-tolerant gadgets, and
intuitively we expect that logical improvement over this baseline is
necessary for an improvement with gadgets from
Ref.~\onlinecite{Aliferis2006}.

The comparison between a sequence of two \CNOT s and a round of error
correction is motivated by the proposal in
Ref.~\onlinecite{Gottesman2016}. Instead of benchmarking the entire
circuit family implied by \cref{fig:1-round-logical-circuit}, we
choose to compare against circuits with two \CNOT s since these are the
physical gates with highest error rates in the physical system.
Although in a fully scalable sub-threshold setting all logical gates
can be made to improve upon their physical counterparts (by going to
sufficiently high distance), the first bar that an experimental
demonstration must meet is the demonstration of {\em some} non-trivial
logical circuit that improves over its physical counterpart, which is
our aim here.

\subsection{Compiler optimizations}

Time spent shuttling and cooling ions can lead to significant memory
error, especially in error correction circuits that tend to leave some
qubits idling for extended times~\cite{RyanAnderson2021}.
Effective memory errors in H2 were previously reported in
Ref.~\onlinecite{Moses2023} for certain types of circuits with random
connectivity, but the circuits used in this study are highly
structured, leading to more coherent accumulation of noise, which is
especially difficult to model accurately. To mitigate this problem,
for the Carbon code experiments, we (1) modified the
optimization of transport operations in the compiler and (2)
incorporated dynamical decoupling pulses into the compiled circuits.

For the first optimization, we used a new cost function to find the
optimal qubit assignments and transport routing, and we allowed the
optimizer to run longer than it normally would during commercial
operations.  The new transport optimization cost function was created
to better account for the time spent during gating relative to the
time spent re-arranging qubits. The net result of this change was that
more gates could be done in parallel reducing the number of gating
steps at the expense of slightly less optimal re-arrangement.  This
optimization resulted in an approximate $15\%$ reduction in the
syndrome extraction time, significantly reducing memory errors. For
the second optimization, the output of the compiler above, which
contains the scheduling information for transport and quantum
operations, was altered by inserting dynamical decoupling pulses into
the schedule opportunistically to ensure no reschedule of operations
or additional transport was needed. Both of these techniques are in
experimental phases and not yet available to most H-series
users. However, especially for the circuits presented in
\cref{sec:repeated-error-correction}, the observed post-selection
rejection rates for the complete circuits improved by $2\times$.

\subsection{Experimental results}

\begin{figure}
  \includegraphics[width=\columnwidth]{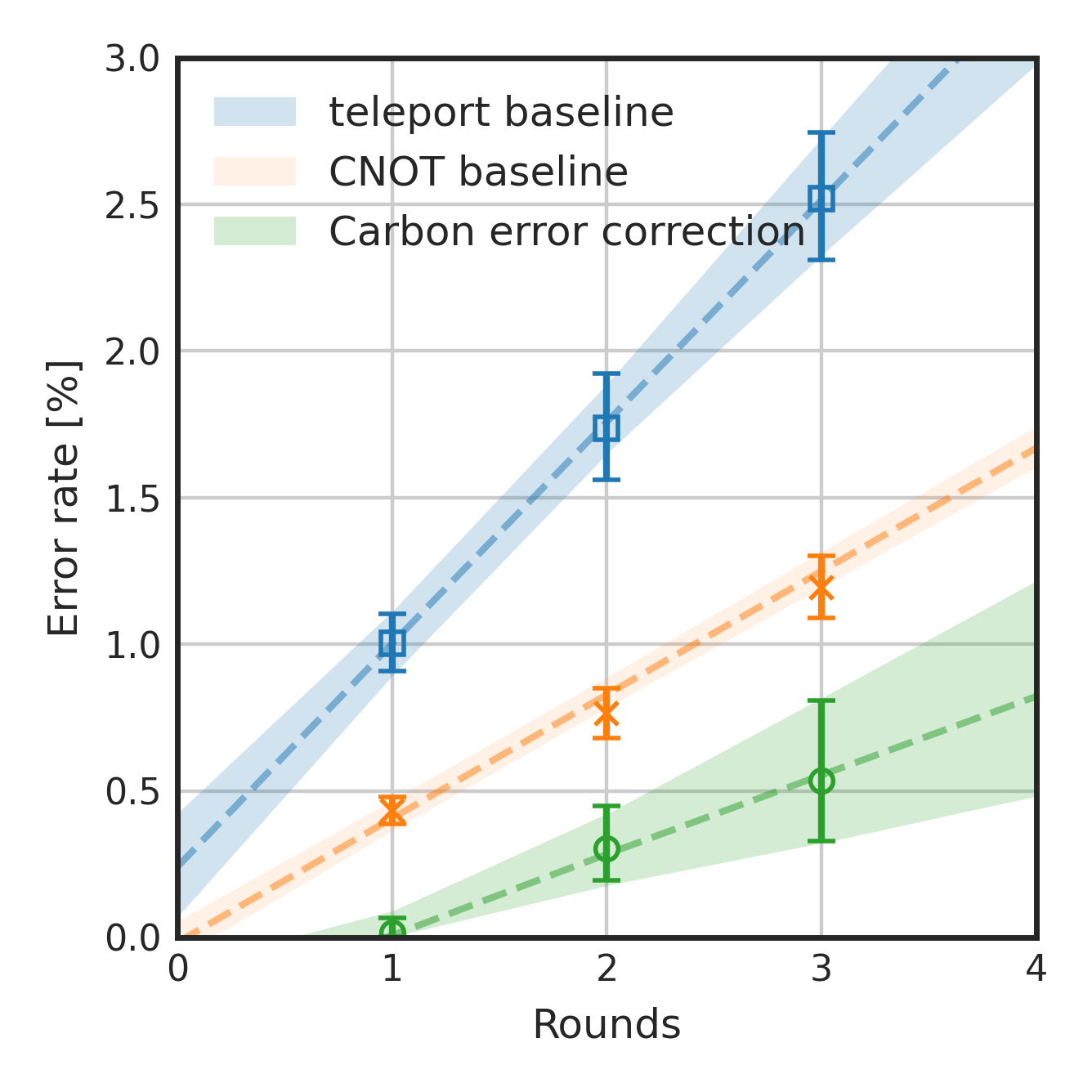}
\caption{Observed error rate for circuits with 1 to 3 rounds of error
  correction with the $[[12,2,4]]$ Carbon code (green circles) and
  physical baselines (blue squares for pairs of 1-bit teleportations,
  and orange x's for pairs of \CNOT s). Results are offset along
  the x-axis for clarity. Linear fits are obtained by
  maximum-likelihood estimation (see \cref{app:statistical-analysis}
  for details).
  \label{fig:error-correction-trend}}
\end{figure}

We observe a gain of more than an order of magnitude for a single
round of error correction, as evidenced by a logical error rate of
$0.017\%_{-0.001\%}^{+0.050\%}$, while the two physical baselines yield
error rates of $1.0\%\pm0.1\%$ (for a sequence of two 1-bit
teleportations) and $0.43\%\pm0.05\%$ (a sequence of two \CNOT s).  For
two rounds of error correction, the logical error rates increase to
$0.3\%\pm0.1\%$, while the two physical baselines' error rates increase
to $1.7\%_{-0.1\%}^{+0.2\%}$ and $0.76\%_{-0.08\%}^{+0.09\%}$ respectively,
which represents a logical gain of more than 2 over the physical
baselines.  For three rounds of error correction the logical error
rate increases to $0.5\%_{-0.2\%}^{+0.3\%}$, which still represents a gain
of more than 2 over the physical baselines of $2.5\%\pm0.2\%$ and
$1.2\%\pm0.1\%$.  Details for each experiment and associated statistical
uncertainties can be found in \cref{tab:3-EC-carbon-results}, and the
results are illustrated in \cref{fig:error-correction-trend}.

\begin{figure}
  \includegraphics[width=\columnwidth]{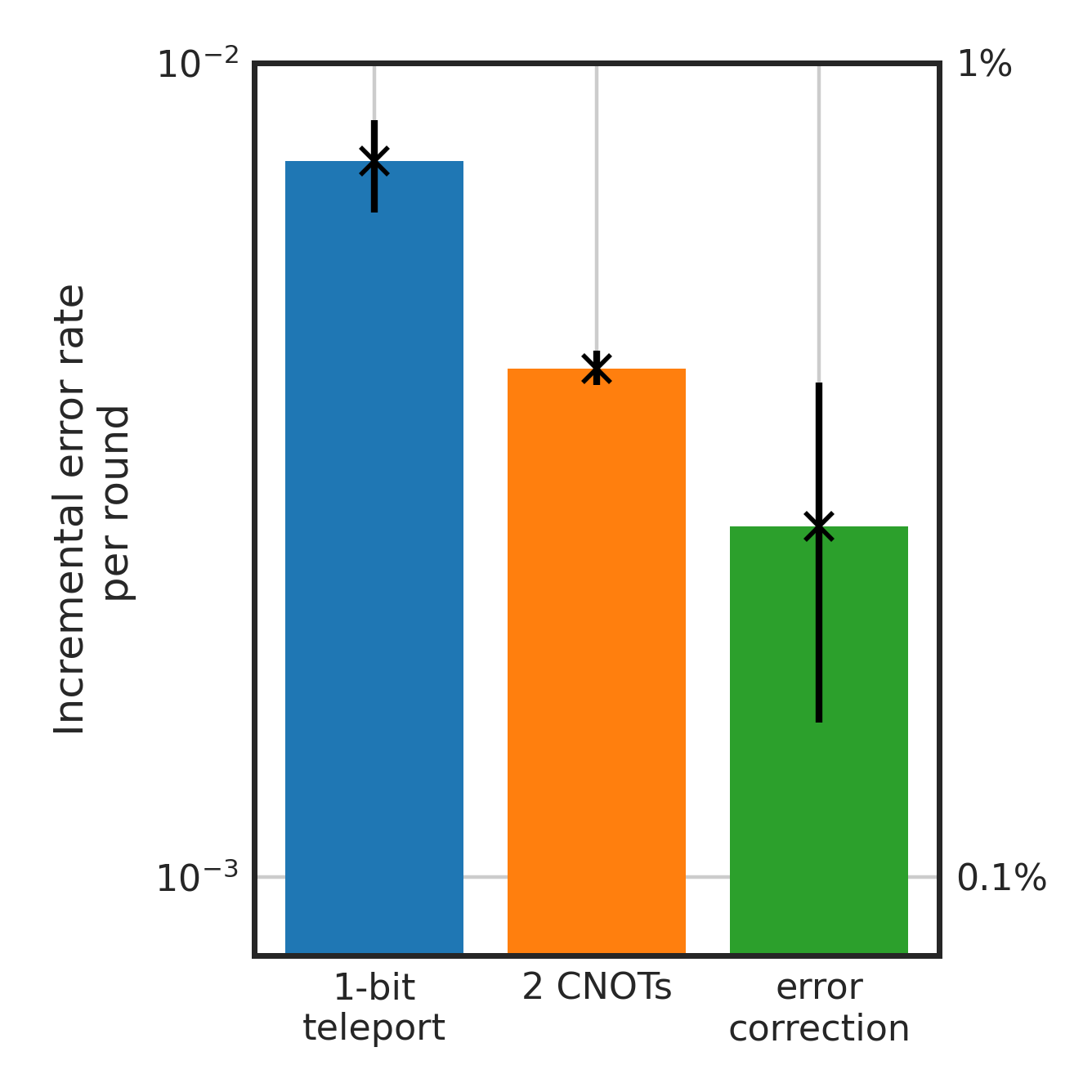}
  \caption{Incremental error rate obtained from maximum likelihood
    linear fits to the data in \cref{fig:error-correction-trend} (see
    \cref{app:statistical-analysis} for details).
  \label{fig:error-correction-trend-slope}}
\end{figure}

Although the gains over the physical error rates appear to behave
non-linearly, the trend for the observed error rate per round of error
correction is consistent with the expected linear behavior for a small
number of repetitions, as illustrated in
\cref{fig:error-correction-trend-slope}. The maximum-likelihood fits
yield error rates per round of the 1-bit teleportation and the
\CNOT\ baseline of $0.7\%\pm0.1\%$ and $0.42\%\pm0.02\%$
respectively, while the fit for each round of error correction points
to $0.3\%\pm0.1\%$---indicating that the differences in the
error rates per round are not statistically significant, largely due
to the uncertainty in the firt for error correction.

The apparent non-linear trend in the gain is an artifact of the
different $y$-intercepts for the linear trends of each of the sets of
experiments. For the physical experiments, the $y$-intercept is set by
the physical state-preparation and measurement (SPAM) error rate. This
is also the case for the logical circuits, but details of the circuit
implementation leads to a negative $y$-intercept despite positive SPAM
error rates. This can be explained by considering optimizations in
scheduling of the operations in the experimental device. In
particular, since memory and transport errors are the dominating
source of error in the experiments (see
\cref{sec:hw-compiler-details}), we estimate the optimized compiled
circuit to have roughly $2r-1$ periods of significant waiting and
transport for $r$ rounds of error correction (due to details of the
initial state preparation and final state measurement), so the logical
error rate for $r$ rounds is estimated to be roughly
\begin{align}
  p_L (2r-1) + p_{{\rm SPAM},L} = 2~p_L~r + (p_{{\rm SPAM},L}-p_L),
\end{align}
where $p_L$ is the logical error rate per memory/transport period, and
$p_{{\rm SPAM},L}$ is the logical SPAM error rate. Since we expect
that the logical error rate per memory/transport period is larger than
the logical SPAM error rate, the intercept for the overall circuit
logical error rate should be negative, agreeing with our experimental
observations.

Surprisingly, there is no significant difference in the error rate per
round of the two physical baselines, although absolute error rates are
slightly more favorable for the physical \CNOT\ baseline, likely due
to it having fewer physical measurements.

The pre- and post-rejection rates also serve to inform whether the
error correction circuits are behaving as expected.  In particular
we expect the pre- and post-rejection rates to be linear in
the number of repetitions (to leading order), and the experimental data
is highly consistent with that prediction (see \cref{sec:rejection-rates}).

\begin{table*}[t]
  \small \centering \renewcommand{\arraystretch}{1.5}
\begin{tabular}{rlrrrrrlr}
    \toprule
&     & runs   & pre-accepted & post-accepted & corrections & errors & error rate & gain \\
%% 1 round %%%%%%%%%%%%%%%%%%%%%%%%%%%%%%%%%%%%%%%%%%%%%%%%%%%%%%%%%%%%%%%%%%%%%%%%%%%%%%%%%%%%%%%%%%%%%%%%%%
\hline
\parbox[t]{2mm}{\multirow{3}{*}{\rotatebox[origin=c]{90}{1 round}}}
&
unencoded baseline (2 teleports)
& \num{40000} &       ---    & ---           & ---         &    401 & $1.0\%_{-0.1\%}^{+0.1\%}$     & ---\\
&
unencoded baseline (2 \CNOT s)
& \num{80000} &       ---    & ---           & ---         &    345 & $0.43\%_{-0.05\%}^{+0.05\%}$  & ---\\
&
encoded, pre-
and
post-selection
& \num{10000} &  \num{7094}  & \num{7008}    & \num{1418}  &     1 & $0.017\%_{-0.001\%}^{+0.050\%}$& 25--58 \\
%% 2 rounds %%%%%%%%%%%%%%%%%%%%%%%%%%%%%%%%%%%%%%%%%%%%%%%%%%%%%%%%%%%%%%%%%%%%%%%%%%%%%%%%%%%%%%%%%%%%%%%%%%
\hline
\parbox[t]{2mm}{\multirow{3}{*}{\rotatebox[origin=c]{90}{2 rounds}}}
&
unencoded baseline (4 teleports)
& \num{20000} &       ---    & ---           & ---         &    347 & $1.7\%_{-0.1\%}^{+0.2\%}$     & ---\\
&
unencoded baseline (4 \CNOT s)
& \num{40000} &       ---    & ---           & ---         &    305 & $0.76\%_{-0.08\%}^{+0.09\%}$  & ---\\
&
encoded, pre-
and
post-selection
& \num{14548} &   \num{7528} & \num{7294}    & \num{2892}    &     22 & $0.3\%_{-0.1\%}^{+0.1\%}$     & 2.5--5 \\
%% 3 rounds %%%%%%%%%%%%%%%%%%%%%%%%%%%%%%%%%%%%%%%%%%%%%%%%%%%%%%%%%%%%%%%%%%%%%%%%%%%%%%%%%%%%%%%%%%%%%%%%%%
\hline
\parbox[t]{2mm}{\multirow{3}{*}{\rotatebox[origin=c]{90}{3 rounds}}}
&
unencoded baseline (6 teleports)
& \num{20000} &       ---    & ---           & ---         &   504 & $2.5\%_{-0.2\%}^{+0.2\%}$     & ---\\
&
unencoded baseline (6 \CNOT s)
& \num{40000} &       ---    & ---           & ---         &   477 & $1.2\%_{-0.1\%}^{+0.1\%}$     & ---\\
&
encoded, pre-
and
post-selection
& \num{10000} &   \num{4029} & \num{3588}    & \num{1896}  &     19 & $0.5\%_{-0.2\%}^{+0.3\%}$    & 2.4--5 \\
    \bottomrule
\end{tabular}
\caption{Summary of experimental results for 1 to 3 rounds of error
  correction via 1-bit teleportations using the $[[12,2,4]]$ Carbon
  code.\label{tab:3-EC-carbon-results}}
\end{table*}

% section: Summary
\section{Summary and Outlook}

We have demonstrated several fault-tolerant circuits outperforming
their physical counterparts in a state-of-the-art QCCD trapped-ion
system. Using different codes and protocols, we demonstrated error
rates 4.7 to 800 times lower than the physical error rates in Bell
correlation experiments with the $[[7,1,3]]$ and the $[[12,2,4]]$
codes. Moreover, we demonstrated up to ten rounds of error correction
for the $[[12,2,4]]$ quantum code by combining error correction and
error detection, with logical error rates well below a physical
baseline circuit of two \CNOT s per round.  With these results, we
have demonstrated that current quantum processors are already able to
reduce error rates in small circuits through quantum error
correction. Future work will focus on extending these results to a
richer set of fault-tolerant logical gadgets, and to ultimately enable
universal fault-tolerant quantum computation while continuing to lower
the achievable logical error rates. A significant milestone will be to
demonstrate a universal family of quantum circuits with logical error
rates significantly below the corresponding physical error rates.

\begin{acknowledgments}
  We thank the broader quantum teams at Microsoft Azure Quantum and
  Quantinuum for many helpful discussions.  We thank Jeongwan Haah for
  insightful discussions and helpful feedback, 
  and Jonhas Colina for help with developing
  the compiler. A special thanks to Z. Alam and J. Strabley for making
  this collaboration possible.
\end{acknowledgments}

\bibliography{references}

\appendix

\section{Carbon code from Knill's C4/C6 scheme\label{app:carbon}}

\subsection{Carbon definition and background}

In 2005, Knill proposed a fault tolerance scheme based on
concatenation of four-qubit ($C_4$) and six-qubit ($C_6$)
codes~\cite{Knill2005}.  This scheme boasts a circuit noise threshold
of $3\%$, arguably the highest known, but also incurs 
large space and time overheads.
Here, we describe an adaptation of Knill's proposal at the
first level of concatenation that substantially reduces the space and time overheads.
At this first level, the concatenation of $C_4$ into $C_6$ yields a $[[12,2,4]]$ code, 
which we nickname Carbon ($C_{12}$). 
The stabilizers of the code are given in \cref{tab:c12-stabilizers}.

\begin{table}
    \centering
    \begin{tabular}{ccccccccccccc}
       0 & 1 & 2 & 3 & 4 & 5 & 6 & 7 & 8 & 9 & 10 & 11 &\\
       \hline
       $X$& $X$ & $X$ & $X$ & $\cdot$ & $\cdot$ & $\cdot$ & $\cdot$ & $\cdot$ & $\cdot$ & $\cdot$ & $\cdot$ &\\
       $Z$& $Z$ & $Z$ & $Z$ & $\cdot$ & $\cdot$ & $\cdot$ & $\cdot$ & $\cdot$ & $\cdot$ & $\cdot$ & $\cdot$ &\\
        $\cdot$ & $\cdot$ & $\cdot$ & $\cdot$ & $X$ & $X$ & $X$ & $X$ & $\cdot$ & $\cdot$& $\cdot$&  $\cdot$ &\\
        $\cdot$ & $\cdot$ & $\cdot$ & $\cdot$ & $Z$ & $Z$ & $Z$ & $Z$ & $\cdot$ & $\cdot$& $\cdot$&  $\cdot$ &\\
        $\cdot$ & $\cdot$ & $\cdot$ & $\cdot$ & $\cdot$ & $\cdot$ & $\cdot$ & $\cdot$ & $X$ & $X$ & $X$ & $X$ &\\
        $\cdot$ & $\cdot$ & $\cdot$ & $\cdot$ & $\cdot$ & $\cdot$ & $\cdot$ & $\cdot$ & $Z$ & $Z$ & $Z$ & $Z$ &\\
        $X$ & $X$ & $\cdot$ & $\cdot$ & $\cdot$ & $X$ & $\cdot$ & $X$ & $X$ & $\cdot$ & $\cdot$ & $X$ &\\
        $X$ & $\cdot$ & $\cdot$ & $X$ & $X$ & $X$ & $\cdot$ & $\cdot$ & $\cdot$ & $X$ & $\cdot$ & $X$ &\\
        $Z$ & $\cdot$ & $Z$ & $\cdot$ & $\cdot$ & $\cdot$ & $Z$ & $Z$ & $Z$ & $\cdot$ & $\cdot$ & $Z$ &\\
        $Z$ & $\cdot$ & $\cdot$ & $Z$ & $Z$ & $\cdot$ & $Z$ & $\cdot$ & $\cdot$ & $\cdot$ & $Z$ & $Z$ &\\
        \hline
        $X$ & $\cdot$ & $\cdot$ & $X$ & $\cdot$ & $\cdot$ & $\cdot$ & $\cdot$ & $\cdot$ & $\cdot$ & $X$ & $X$ & $=X_0$\\
        $Z$ & $\cdot$ & $\cdot$ & $Z$ & $\cdot$ & $\cdot$ & $\cdot$ & $\cdot$ & $\cdot$ & $Z$ & $\cdot$ & $Z$ & $=Z_0$\\
        $\cdot$ & $X$ & $\cdot$ & $X$ & $\cdot$ & $\cdot$ & $\cdot$ & $\cdot$ & $\cdot$ & $X$ & $X$ & $\cdot$ & $=X_1$\\
        $Z$ & $Z$ & $\cdot$ & $\cdot$ & $\cdot$ & $\cdot$ & $\cdot$ & $\cdot$ & $\cdot$ & $Z$ & $Z$ & $\cdot$ & $=Z_1$\\
    \end{tabular}
    \caption{Carbon stabilizers.}
    \label{tab:c12-stabilizers}
\end{table}

The $C_4$ code and its subsystem variants have been widely
studied~\cite{Vaidman1996,Grassl1997,Bacon2006,Knill2003,Knill2005,Gottesman2016}
and used in proposals for magic state
distillation~\cite{Meier2013,DuclosCianci2015,Campbell2016} as well as
fault-tolerant computation with linear qubit
arrays~\cite{Jones2018,Stephens2009}.  The $C_6$ code has been
proposed for use in magic state distillation~\cite{Jones2013}, but is
much less widely studied.

Knill used teleportation as a means of error correction, this scheme
achieves a remarkably high threshold of $3\%$ error per gate.  A
substantial drawback to Knill's proposal is the qubit resource
overhead.  To execute an algorithm with circuit volume of $10^8$ with
a physical CNOT error rate of $0.1\%$, Knill estimates a qubit
overhead of $10^5$ \emph{per logical gate}.  There are a few reasons
for the hefty overhead.  First, the family of $C_4/C_6$ codes have
even distance. There are some errors that can be detected but not
corrected.  In those situations, the multiple copies of circuits may
be required in order to successfully execute a gate with high
probability.  Second, a key element of Knill's proposal is to perform
error correction by teleportation.  However, the circuitry that Knill
used was not particularly efficient---Knill's $C_4$
preparation circuit requires eight qubits and eight CNOTs, whereas
fault tolerant preparation circuits with five qubits and five CNOTs
are now known~\cite{Gottesman2016}.  Additionally, Knill did not
consider optimizations across levels of concatenation.

Our adaptation benefits from treating the Carbon code monolithically
rather than as a concatenation---making the circuits and the number of
required ancillas much smaller.  We leverage several modern fault
tolerance innovations developed since Knill's original proposal.  The
resulting circuits for preparing $|00\rangle$ require $19$ qubits and
$37$ CNOTs, whereas Knill's preparation requires $144$ qubits and
$216$ CNOTs~\footnote{The number of qubits depends on the amount of
parallelism. Here we report the number of qubits required in order to
achieve maximum parallelism.}  Our circuits for Bell state preparation
require as few as 36 qubits.  The reduction in
circuit size directly improves acceptance rates, which reduces the
expected number of copies required for success and further reduces
qubit overhead.

\subsection{Fault tolerance with Carbon}

Like Knill, we propose using teleportation to detect and correct
errors and also for implementing several logical operations.  We
describe circuits both for Bell-state (2-bit) teleportation, and for
1-bit teleportation as described in the main body.
The $H\otimes H$ and inter-block CNOT$\otimes$CNOT gates are both
transversal for Carbon, up to permutations.  CNOT and SWAP between
logical qubits within a block can be done by permutation.  Selective
gates, $I\otimes H$ and $I\otimes S$, can be executed by preparing and
teleporting appropriate stabilizer states.  Non-Clifford operations can
be implemented by standard magic state distillation techniques such
as~\cite{Bravyi2005,Jones2013,Meier2013}. 

We limit our focus here to only those operations necessary for
the error correction and Bell state demonstrations described in the main
text, primarily $\ket{00}$ and $\ket{++}$ preparation.
In all cases, circuits for Carbon are designed in order to meet the following two conditions.
\begin{enumerate}
\item A logical error occurs only if there are at least three circuit faults.
\item Post-rejection occurs only if there are at least two circuit faults.
\end{enumerate}

\subsubsection{\texorpdfstring{$\ket{00}$}{00} and \texorpdfstring{$\left|++\right\rangle$}{++} states}

\begin{figure}
  \centering
  \includegraphics[width=\columnwidth]{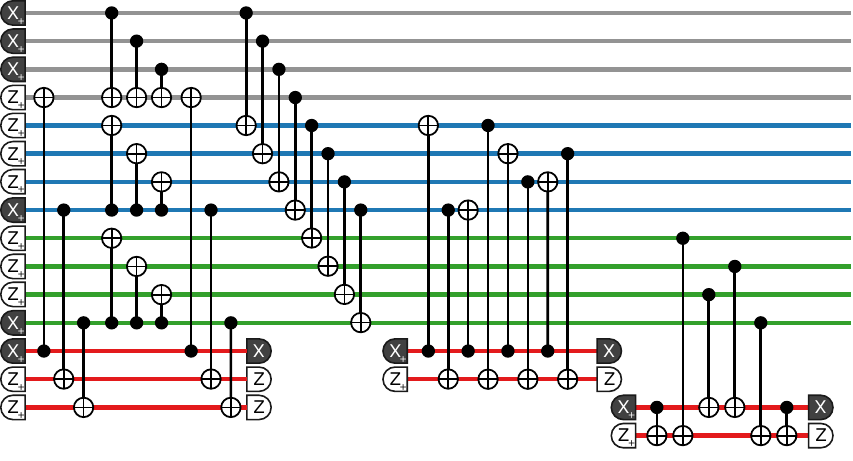}
  \caption{Carbon $\ket{00}$ preparation circuit.  The circuit
    consists of three parts.  First three $C_4$ blocks are prepared,
    indicated by the gray, blue and green qubit lines.  Then CNOTs are
    applied transversally between $C_4$ blocks in order to encode into
    Carbon.  Then three weight-four stabilizers are measured to
    check for high weight errors.  The $C_4$ preparations include
    flags, two of which are then combined with subsequent stabilizer
    measurements in order to reduce circuit size.  Following the CNOT
    gates, permutation $(0~3~1~2)$ is applied to the middle four
    qubits (blue) and permutation $(0~2~3~1)$ is applied to the
    bottom four qubits (green).}
  \label{fig:c12-00}
\end{figure}

A circuit for preparing logical $\ket{00}$ is illustrated in
\cref{fig:c12-00}.  Logical $\ket{++}$ can be prepared by appending
transversal Hadamard and approprate qubit relabeling.  Therefore, the
$\ket{00}$ circuit is the foundation for all of our teleportation
circuits.

In the absence of errors the measurements in \cref{fig:c12-00} should
all be $+1$.  Any $-1$ outcome indicates an error and results in
pre-rejection of the state.  That is, the state is thrown away prior
to interaction with any other encoded blocks.  Conditioned on $+1$
outcomes, the output is ``fault tolerant'' in the following sense. For
any single fault in the circuit, the residual error has CSS weight at
most one, modulo stabilizers.
CSS weight is defined as maximum Hamming weight of the $X$ and $Z$ 
bitstrings in the usual symplectic representation.
For any two faults in the circuit, the
residual error has CSS weight at most two \emph{or} the residual error
has a Carbon syndrome distinct from the syndrome of any CSS
weight-one error (so that it can be post-rejected).

This circuit was designed by adapting and combining several existing
circuits in the literature.  First, Knill's circuit for preparing
$\ket{00}$ for $C_4/C_6$ was stripped of all error detection
circuitry, leaving only the $C_4$ preparations and transversal CNOTs
between $C_4$ blocks.  Next, flags were added to the $C_4$
preparations in order to detect weight-two
errors~\cite{Chao2018}. (There is just a single weight-two error per
preparation.)  Note that other similar $C_4$ preparation circuits
would suffice, such as~\cite{Gottesman2016}.  Then additional
stabilizer measurements were added to detect weight-two errors from
the transversal CNOTs.  The $X^{\otimes 4}$ and $Z^{\otimes 4}$
measurements on the middle four qubits use a circuit proposed by
Reichardt~\cite{Reichardt2020}.  The remaining $Z^{\otimes 4}$
measurement is a flag circuit modified to improve the
depth~\cite{Chao2018}.  

The flag qubits used to detect errors in the $C_4$ preparations midway
through the preparation are
noteworthy.  An alternative strategy is first prepare $\ket{00}$
with a unitary circuit, and postpone flag qubits and measurements until
after the unitary.  This strategy is initially appealing because it is
relatively easy to find unitary preparation circuits and, given a
unitary preparation, it is easy to find sets of stabilizers that
detect correlated errors.  However, using flags early on in the
circuit has two benefits.  First, they can detect high-weight but
manageable errors before they become even higher weight.  A single
fault in the middle $C_4$ preparation, for example, can cause a
weight-two error $X_{7,8}$.  The flag catches this
error before it can subsequently spread to $X_{7,8,11,12}$ when it
would be harder to manage.  Second, fault tolerant detection circuitry
is much simpler to construct.  The $C_4$ flag gadgets don't themselves
cause any high weight errors.  Contrast that with the alternative of
measuring a weight-six stabilizer at the end of the circuit, which
might require several additional flags.

A few variants of \cref{fig:c12-00} may be useful.  Two of the flags
in the circuit can be combined with stabilizer measurements in order to
reduce circuit size and measurement count.  This may be beneficial if
mid-circuit measurements are expensive.  Another
variant is to postpone the stabilizer measurements, as shown in
\cref{fig:c12-bell}.  This can be useful when using $\ket{00}$ as
a component for a larger state.  This approach is used, for example,
to prepare Bell states, as described below.

\subsubsection{Bell states}

\begin{figure*}
  \centering
  \includegraphics[width=\textwidth]{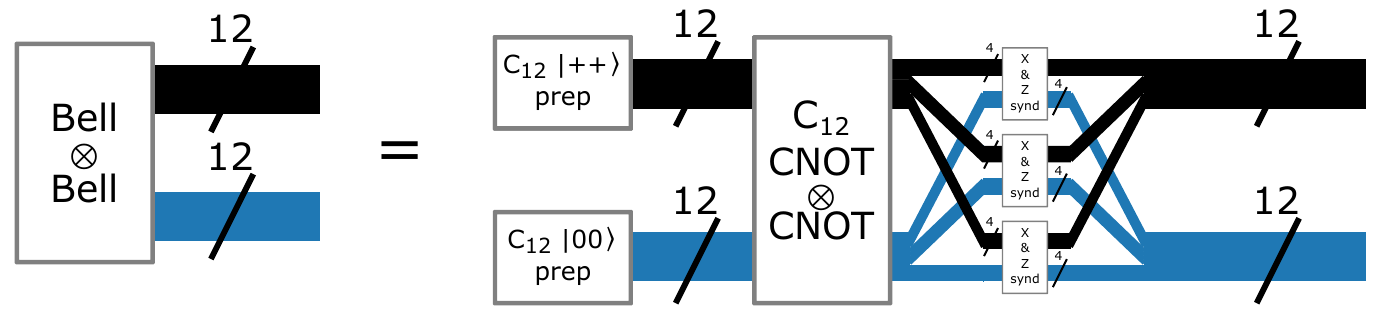}
  \caption{Carbon Bell state preparation circuit.  This circuit
    prepares two Bell states across two Carbon blocks.  One Carbon
    block is prepared as $\ket{++}$ and the other as $\ket{00}$.
    Note that the weight-four measurements illustrated in \cref{fig:c12-00}
    are not necessary here.
    Then, corresponding $C_4$ blocks are paired up, followed by
    transversal CNOTs across each pair. Following the transversal
    CNOTs, $X^{\otimes 4}$ is measured on the top block of each pair
    and $Z^{\otimes 4}$ is measured on the bottom block of each pair.
    If any of the measurements have value $-1$, then the state is
    discarded.  The $X^{\otimes 4}$, $Z^{\otimes 4}$ measurement
    circuit is given in~\cref{fig:c4-error-detect}.  
}
  \label{fig:c12-bell}
\end{figure*}

\begin{figure*}
  \centering
  \includegraphics[width=\textwidth]{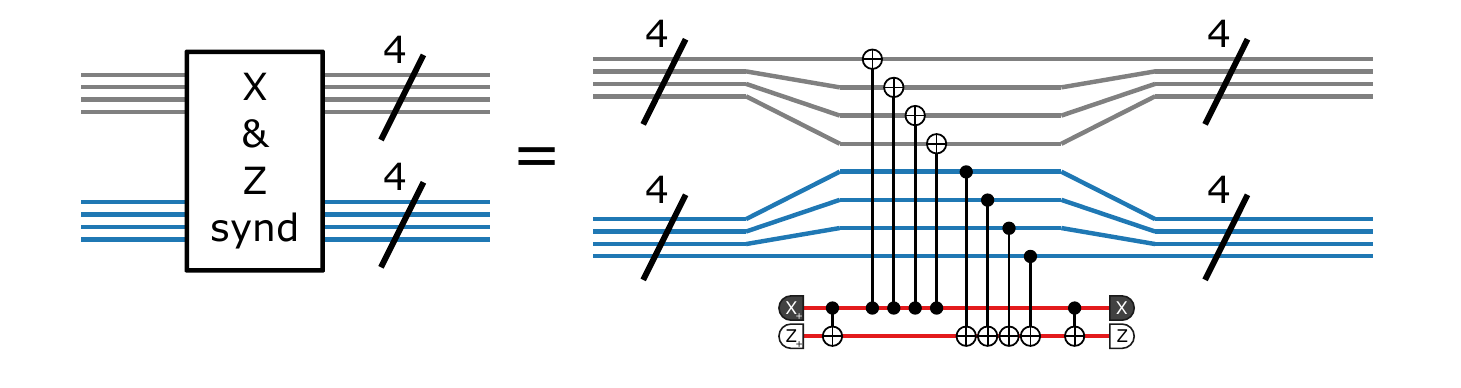}
  \caption{Simultaneous measurement of $X^{\otimes 4}$ and $Z^{\otimes
      4}$ on a pair of $C_4$ blocks.  This circuit is similar to that
    of~\cite{Jones2018}, but measures operators on two different code
    blocks.  This circuit is fault tolerant in the sense that any
    single fault is either detected (by $-1$ outcome of a measurement)
    or results in an error of at most weight one (modulo $X^{\otimes
      4}$ and $Z^{\otimes 4}$ on the data qubits.  CNOTs on data
    qubits are shown sequentially for illustration, but may be
    re-ordered arbitrarily, or parallelized as desired.}
  \label{fig:c4-error-detect}
\end{figure*}

Since CNOT is transversal for Carbon, a pair of Bell states across
two blocks can be prepared in the obvious way by first fault
tolerantly preparing $\ket{++}$ and $\ket{00}$ as described above,
and then applying transversal CNOTs across the two blocks.  An
alternative approach is illustrated in \cref{fig:c12-bell}.  This
approach also applies transversal CNOTs across $\ket{++}$ and
$\ket{00}$ blocks. But now stabilizer measurements are postponed
until \emph{after} the transversal CNOTs.  Doing so allows for
detecting additional errors on the transversal CNOTs, which could be
an advantage depending on noise model and architecture.  It can also
be an advantage when preparing other entangled resource states for
gate teleportation.  The
detection circuitry is shown in greater detail in
\cref{fig:c4-error-detect}.

\subsubsection{Error correction}

Given a Bell state, as prepared above, fault tolerant error correction
can be implemented by teleportation, as described by Knill.  Half of
the Bell states, one of the Carbon blocks, interacts with an
input Carbon ``data'' block.  Then both the data and the Bell state
half are measured.  This has the effect of teleporting the data block
into the remaining half of the Bell state \emph{and} measuring a
Carbon syndrome.  The syndrome can then be used to detect or correct
errors on the (teleported) data.

The syndrome can be decoded with a specially modified lookup decoder.
A typical lookup decoder for an $n$-qubit code is constructed by
enumerating over all $n$-qubit Pauli errors and computing each
syndrome.  But Carbon is a distance four code.  Some distinct
weight-two Pauli errors will generate the same syndrome and are
therefore not correctable.  To achieve an effective distance of five,
so that logical errors occur only if there are at least three circuit
faults, the lookup decoder can be modified so that uncorrectable
syndromes are removed from the table.  If such a syndrome occurs at
runtime, then we ``post-reject'' by throwing those computations away.

In~\cite{Knill2005}, Knill uses a phenomenological error model, in
which errors occur independently on qubits, in order to determine
which syndromes to remove from the lookup table.  That model is not
sufficient for our circuits.  We allow pairs of faults to pass
pre-selection if the syndrome of the residual error is distinct from
that of all CSS-weight one Pauli errors.  This yields efficient
circuits, but can also allow for what appear to be correctable
weight-two errors based on syndrome, but are actually errors of higher
weight.  

One way to account for this is by tailoring the lookup table
according to the state preparation circuits for $\ket{00}$ and
$\ket{++}$.  The table is constructed by enumerating over all single
and double faults within the circuits. If, for a given syndrome, the
residual error of the faults always matches the naive lookup table,
then the syndrome is kept in the table.  Otherwise, the syndrome is
thrown out.

We, however, adopt a simpler stategy:
start with the naive Carbon lookup decoder, but
keep only the syndromes that are consistent with CSS weight-one Pauli
errors. This option is more aggressive in that it throws away some
error configurations that might be correctable, but it maintains a
$O(p^2)$ post-rejection rate. We find that it impacts the post-rejection
rate only marginally, while substantially improving logical
error rates.

\subsection{Scaling to larger sizes}

In Knill's original proposal Carbon ($C_4/C_6$) was concatenated
repeatedly with $C_6$ in order to increase the code distance.  Knill
suggested concatenating with other codes once the effective error rate
is sufficiently low.  The same approach is compatible with our scheme
for Carbon.  Our circuit optimizations do not immediately apply to
further concatenations to $C_6$, as they do not respect the boundaries
between $C_4$ and $C_6$, but Knill's circuit constructions for $C_6$
can be applied without modification.  Swapping out Knill's $C_6$
encoding circuits for a flag-based stabilizer measurement approach is
likely not beneficial because it would require selective CNOT gates,
which are somewhat complex for these codes (require specialized state
preparation and teleportation). Other optimizations may still be
possible. For example, Knill uses Bell pairs followed by lower level
error correction to encode $C_6$. When the lower level is Carbon,
the error correction can be elided in favor of using
\cref{fig:c12-bell}.  This saves a factor of three in qubit overhead.

Concatenation has the benefit of blunting the impact of
post-selection.  A syndrome that induces post-rejection in Carbon
can instead be corrected as an erasure at higher levels of encoding.
If error correction by teleportation is also used at higher levels,
then Carbon gate gadgets will be entirely (or almost entirely)
contained in state preparation circuits for those higher levels.
Post-selection in Carbon is thereby converted into
\emph{pre}-selection for the larger code.

\section{Statistical analysis and error bars\label{app:statistical-analysis}}

Error bars for failure rates are computed by assuming each
experimental run of a particular circuit is an independent and
identically distributed Bernoulli trial, so that the total count of
failures given the number of trials is binomially distributed. Since the
binomial distribution has the beta distribution as a conjugate prior
distribution of the binomial parameter $0\le p\le 1$, we estimate the
error bars by updating a beta prior with the outcome of the
experiments.

More precisely, taking the prior distribution
\begin{align}
  p \sim \Beta(\alpha, \beta),
\end{align}
the posterior distribution after running $N$ experiments with $F$
failures is
\begin{align}
  p \sim \Beta(\alpha+F, \beta+N-F),
\end{align}
by applying Bayes' theorem to update our belief about the parameter
$p$.  We take $(\alpha,\beta)=(\frac{1}{2},\frac{1}{2})$, known as
{\em Jeffreys' prior}.

Throughout, the point estimates we report are the medians of the
posteriors, while the error bars correspond to the $2.5\%$ and the
$97.5\%$ of the posteriors, yielding a $95\%$ credible
interval. Following standard practice~\cite{Taylor1996}, we set
significant digits for these estimate based on the most significant
digit of the lower bound of the credible interval.

These estimates are conservative, in the sense that they are generally
biased away from 0 and 1. Notably, even if no failures are detected,
the median of the posterior will be non-zero, so that our point
estimates for the error probability are never 0. The upper bound of
the credible interval is marginally more conservative than the maximum
risk $95\%$ confidence ``rule of 3'' estimate used in medical and
engineering fields~\cite{Hanley1983,Eypasch1995}.

The linear fits depicted in \cref{fig:error-correction-trend}, are
estimated by maximum likelihood fitting of the linear model
parameters. The likelihood is given by the joint probability density
function of the observations, which consists of the product of the
beta-distributed posteriors as outlined above. The uncertainty in the
fit parameters is too large for conclusive statements about how the
average accumulated error rates per round compare, although it is
apparent that the gap between them is not large.

\section {Rejection rates in the Carbon code\label{sec:rejection-rates}}

While we are particularly interested in the behavior of logical error
rates for the fault-tolerant circuits, the behavior of the rejection
rates for pre-selection and post-selection also informs us about the
performance of the fault-tolerant circuits.

First, under the assumption of independent errors at each circuit
location, pre-selection rejections are expected to be first-order in
the physical error rates (recall pre-selection is not a barrier to
scalability, since state preparation factories can be used to increase
the probability of successful preparation exponentially close to
1). When error correction and error rejection are combined as we did
in \cref{subsec:carbon-code-bell,sec:repeated-error-correction}, on
the other hand, cause the post-selection rejections rates to be
second-order in the physical error rates, and therefore much smaller.

Second, for small circuits, the rejection rates should increase
roughly linearly with the logical volume of the circuit. In our case,
the logical volume is simply proportional to the number of error
correction rounds.

\begin{figure}
  \includegraphics[width=\columnwidth]{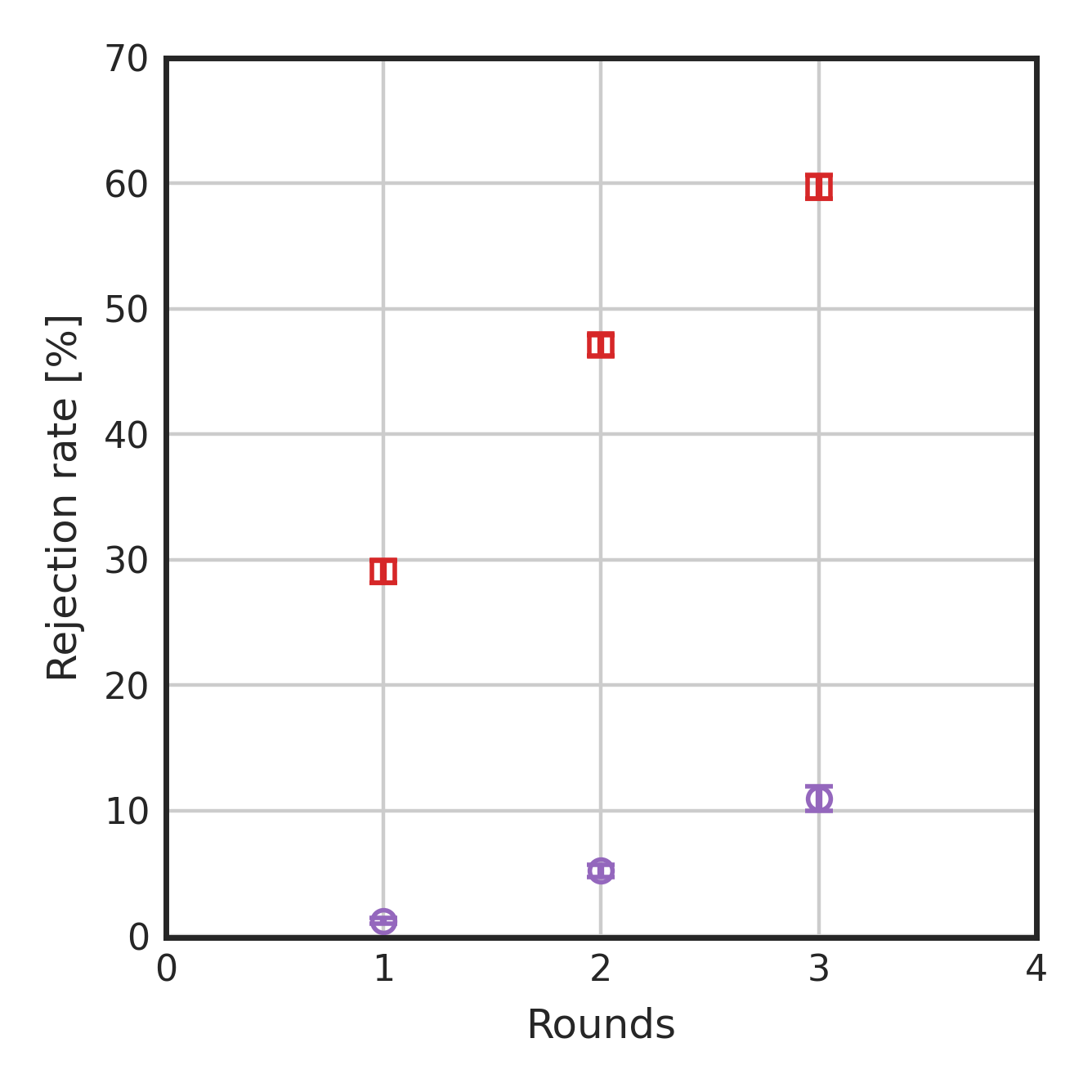}
  \caption{Rejection rates during pre-selection (red squares) and
    post-selection (purple circles) as a function of the number of
    error correction rounds for the $[[12,2,4]]$ Carbon code. In both
    cases the expected linear trend (as a function of number of
    repetitions) is apparent, as is the large separation between
    pre-rejections and post-rejections.\label{fig:rejection-rates}}
\end{figure}

These expected trends are apparent in \cref{fig:rejection-rates},
as is the large separation between the pre- and post-rejection rates,
consistent with what would be expected between first- and second-order
events.

\FloatBarrier
\section{Additional experimental data for the Steane code experiments\label{app:steane-data}}

The Steane code admits transversal $Y \otimes Y$ measurements, allowing a straight-forward measurement of the state fidelity for the Bell state preparation. For a Bell pair, the state fidelity is given by $F_s=(I \otimes I+X \otimes X-Y \otimes Y+Z \otimes Z)/4$. In this section, Table~\ref{Table:BellSteanePhys}, Table~\ref{Table:BellSteaneQEC}, and Table~\ref{Table:BellSteaneQED} provide the state fidelities for the various Steane code experiments and their physical level analogs, and a break-down of the results into the three different Pauli measurements required.

Note that the ``gain'' values in Table~\ref{Table:BellSteaneQEC} and Table~\ref{Table:BellSteaneQED} are calculated by dividing the state fidelity (which is calculated using the $X \otimes X$, $Y \otimes Y$, and $Z \otimes Z$ bases measurements) of the unencoded circuit by the state fidelity of the encoded circuit.

% Physical
 \begin{table}[t]
\begin{tabular}{crl}
\toprule
Pauli           & runs      & error rate \\ 
\hline
$X \otimes X$   & $137,200$   & $0.42\% _{-0.03\%}^{+0.04\%}$         \\
$Y \otimes Y$   & $137,200$   & $0.39\% _{-0.03\%}^{+0.03\%}$     \\ 
$Z \otimes Z$   & $137,200$   & $0.58\% _{-0.04\%}^{+0.04\%}$      \\
\hline
$E_{xz}$        & $274,400$ & $0.50\% _{-0.03\%}^{+0.03\%}$    \\ 
$E_s$           & NA & $0.70\% _{-0.03\%}^{+0.03\%}$    \\ 
\bottomrule
\end{tabular}
\caption{The measured state infidelity $E_s=1-F_s$ of the physical-level Bell preparation gadget and the error rates of individual Pauli operators $X \otimes X$, $Y \otimes Y$, and $Z \otimes Z$. For comparison, the error rate determined from measuring the probability of getting the wrong parity from $X \otimes X$ and $Z \otimes Z$ is given as $E_{xz}$.}
\label{Table:BellSteanePhys}
\end{table}

% Logical - QEC
 \begin{table}[t]
\begin{tabular}{crrlr}
\toprule
Pauli           & runs & pre-accepted  & error rate                & gain \\ 
\hline
$X \otimes X$   & $12,100$ & $9,025$    & $0.10\% _{-0.05\%}^{+0.08\%}$       & $4.2$      \\ 
$Y \otimes Y$   & $12,100$ & $9,082$    & $0.09\% _{-0.05\%}^{+0.08\%}$       & $4.3$      \\ 
$Z \otimes Z$   & $12,100$ & $9,010$    & $0.003\% _{-0.003\%}^{0.025\%}$  & $193$     \\ 
\hline
$E_{xz}$          & $24,200$ & $18,035$ & $0.05 \% _{-0.03\%}^{+0.04\%}$     & $9.8$ \\
$E_s$          & NA & NA & $0.10\% _{-0.04\%}^{+0.06\%}$       & $6.8$ \\ 
\bottomrule
\end{tabular}
\caption{The measured state infidelity $E_s=1-F_s$ of the logical-level Bell preparation gadget utilizing the Steane code where the destructive measurements are analyzed using quantum error correction. The error rates of individual Pauli operators $X \otimes X$, $Y \otimes Y$, and $Z \otimes Z$ are also reported. For comparison, the error rate determined from measuring the probability of getting the wrong parity from $X \otimes X$ and $Z \otimes Z$ is given as $E_{xz}$.}
\label{Table:BellSteaneQEC}
\end{table}

% Logical - QED
 \begin{table}[t]
\begin{tabular}{crrrlr}
\toprule
Pauli           & runs & pre- & post- & error rate & gain\\ 
           &  & accepted & accepted & & \\ 
\hline
$X \otimes X$   & $12,100$ & $9,025$ & $8,688$ & $0.003\% _{-0.003\%}^{+0.026\%}$      &  $140$       \\ 
$Y \otimes Y$   & $12,100$ & $9,082$ & $8,665$ & $0.03\% _{-0.02\%}^{+0.05\%}\%$      & $13$     \\ 
$Z \otimes Z$   & $12,100$ & $9,010$  & $8,701$ & $0.003\% _{-0.003\%}^{+0.026\%}$     & $193$     \\  
\hline
$E_{xz}$          & $24,200$ & $18,035$ & $17,389$ & $0.001 \% _{-0.001\%}^{+0.013\%}$     & $500$ \\ 
$E_s$          & NA & NA & NA & $0.02 \% _{-0.01\%}^{+0.03\%}$     & $35$\\ 
\bottomrule
\end{tabular}
\caption{The measured state infidelity $E_s=1-F_s$ of the logical-level Bell preparation gadget utilizing the Steane code where the destructive measurements are analyzed using QED. The error rates of individual Pauli operators $X \otimes X$, $Y \otimes Y$, and $Z \otimes Z$ are also reported. For comparison, the error rate determined from measuring the probability of getting the wrong parity from $X \otimes X$ and $Z \otimes Z$ is given as $E_{xz}$.}
\label{Table:BellSteaneQED}
\end{table}

\FloatBarrier

\end{document}